\documentclass[8.5pt,twoside,twocolumn]{article}
\usepackage[super,sort&compress,comma]{natbib}
\usepackage{mhchem}
\usepackage{times,mathptmx}
\usepackage{sectsty}
\usepackage{balance}
\usepackage{multirow}
\usepackage{epsfig}
\usepackage{lastpage}
\usepackage[format=plain,justification=raggedright,singlelinecheck=false,font=small,labelfont=bf,labelsep=space]{caption}

\begin{document}

\thispagestyle{plain}

\makeatletter
\def\subsubsection{\@startsection{subsubsection}{3}{10pt}{-1.25ex plus -1ex minus -.1ex}{0ex plus 0ex}{\normalsize\bf}}
\def\paragraph{\@startsection{paragraph}{4}{10pt}{-1.25ex plus -1ex minus -.1ex}{0ex plus 0ex}{\normalsize\textit}}
\renewcommand\@biblabel[1]{#1}
\renewcommand\@makefntext[1]%
{\noindent\makebox[0pt][r]{\@thefnmark\,}#1}
\makeatother
\renewcommand{\figurename}{\small{Fig.}~}
\sectionfont{\large}
\subsectionfont{\normalsize}


\twocolumn[
  \begin{@twocolumnfalse}
\noindent\LARGE{\textbf{Controlling disorder in the ZnGa$_2$O$_4$:Cr$^{3+}$ persistent phosphor by Mg$^{2+}$ substitution}}
\vspace{0.6cm}

\noindent\large{\textbf{Neelima Basavaraju,\textit{$^{a}$}\textit{$^\dag$} Kaustubh R. Priolkar,$^{\ast}$\textit{$^{a}$} Aur\'{e}lie Bessi\`{e}re,\textit{$^{b}$} Suchinder K. Sharma,\textit{$^{b}$}\textit{$^\ddag$} Didier Gourier,\textit{$^{b}$} Laurent Binet,\textit{$^{b}$} Bruno Viana\textit{$^{b}$} and Shuichi Emura\textit{$^{c}$}}}\vspace{0.5cm}

\noindent\textit{\small{\textbf{Received Xth XXXXXXXXXX 20XX, Accepted Xth XXXXXXXXX 20XX\newline
First published on the web Xth XXXXXXXXXX 200X}}}

\noindent \textbf{\small{DOI: 10.1039/b000000x}}
\vspace{0.6cm}

\noindent \normalsize{We have studied in this work the effect of increasing structural disorder on the persistent luminescence of Cr$^{3+}$ doped zinc gallate spinel. This disorder was introduced by progressive substitution of Zn$^{2+}$ by Mg$^{2+}$ ions, and was studied by photoluminescence, X-ray diffraction, extended X-ray absorption fine structure (EXAFS), X-ray absorption near edge structure (XANES) and electron paramagnetic resonance (EPR) spectroscopy. It was found that increasing Mg/Zn substitution decreases the number of Cr$^{3+}$ in undistorted sites, and increases the number of Cr$^{3+}$ with neighbouring antisite defects and with neighbouring Cr$^{3+}$ ions (referred as Cr clusters), which in turn decreases the intensity of persistent luminescence. Both XANES and EPR spectra could be simulated by a linear combination of Cr$^{3+}$ spectra with three types of Cr$^{3+}$ environments. The increasing disorder was found to be correlated with a decrease of the average Cr-O bond length and a decrease of crystal field strength experienced by Cr$^{3+}$ ions.}
\vspace{0.5cm}
 \end{@twocolumnfalse}
  ]



\footnotetext{\textit{$^{a}$Department of Physics, Goa University, Taleigao plateau, Goa 403206, India. Tel: 0832 651 9084; E-mail: krp@unigoa.ac.in}}
\footnotetext{\textit{$^{b}$PSL Research University, Chimie ParisTech - CNRS, Institut de Recherche de Chimie Paris, 75005, Paris, France.}}
\footnotetext{\textit{$^{c}$Institute of Scientific and Industrial Research, Osaka University, 8-1 Mihogaoka, Ibaraki, Osaka 567-0047, Japan.}}
\footnotetext{\textit{$^\dag$Solid State and Structural Chemistry Unit, Indian Institute of Science, Bengaluru 560012, India.}}
\footnotetext{\textit{$^\ddag$Department of Physics, Chalmers University of Technology, SE-412 96, G\"{o}teborg, Sweden.}}

\section{Introduction}

Persistent phosphors which show near infrared persistent luminescence (hereafter referred to as NIRPL) were first demonstrated in 2007 to be used as biomarkers for \textit{in vivo} small animal optical imaging.\cite{1} This technique was illustrated using silicate nanoparticles Ca$_{0.2}$Zn$_{0.9}$Mg$_{0.9}$Si$_{2}$O$_{6}$:Eu$^{2+}$, Dy$^{3+}$, Mn$^{2+}$ (CZMSO) with Mn$^{2+}$ as luminescent ion.\cite{1} Soon to follow in 2011, a novel compound Cr$^{3+}$ doped ZnGa$_{2}$O$_{4}$ with enhanced NIRPL properties was reported as a potential probe for \textit{in vivo} imaging application.\cite{2} The origin of NIRPL was explained to be resulting from Cr$^{3+}$ perturbed by neighbouring antisite defects, which are the defects resulting from exchange in site positions of Zn and Ga ions.\cite{2} The suitability of Cr$^{3+}$ doped ZnGa$_{2}$O$_{4}$ biomarker for \textit{in vivo} imaging of tumors was soon demonstrated in 2014 by Maldiney et al..\cite{3} Zinc gallate proved to be an exciting candidate for the application owing to the fact that its NIRPL could be triggered even with visible light excitation, enabling the multiple excitations of nanoparticles through the animal body, making them a perpetual probe.\cite{3,4} Recently, Cr$^{3+}$ doped zinc gallogermanates have also been shown to present very long NIRPL.\cite{5,6,7,8} Another new compound Cr$^{3+}$ doped MgGa$_{2}$O$_{4}$ was reported in 2013 as a possible candidate for optical imaging, stating the key role of structural inversion in governing the NIRPL.\cite{9} This MgGa$_{2}$O$_{4}$ compound presented a much broader and red shifted NIRPL emission as compared to that of Cr$^{3+}$ doped ZnGa$_{2}$O$_{4}$ compound, which was speculated to be more advantageous in terms of \textit{in vivo} imaging application.\cite{9,10}

In order to improve the NIRPL efficiency of these biomarkers for the application, it is necessary to understand the proper mechanism for the origin of persistent luminescence in these materials. Hence, we have undertaken extensive studies on the Cr$^{3+}$ doped AB$_{2}$O$_{4}$ (A = Zn, Mg and B = Ga, Al) spinels by combined optical and structural characterizations.\cite{11,12,13,14} We have proposed a mechanism for NIRPL in ZnGa$_{2}$O$_{4}$:Cr$^{3+}$ which emphasizes the key role played by Cr$^{3+}$ ions with antisite defects as their first cationic neighbour (known as Cr$_{N2}$ ions).\cite{11} Pairs of antisite defects around Cr$^{3+}$ ion are thought to create local electric field which trigger charge separation and trapping, without change in valence state of Cr$^{3+}$ ion.\cite{11} This mechanism was substantiated by electron paramagnetic resonance (EPR) studies on ZnGa$_{2}$O$_{4}$:Cr$^{3+}$ compounds with varying nominal cationic ratio.\cite{12,15} Further, X-ray absorption measurements were carried out on the Cr$^{3+}$ doped AB$_{2}$O$_{4}$ (A = Zn, Mg and B = Ga, Al) spinel hosts which established the main role of inversion in controlling NIRPL.\cite{13,14} Extended X-ray absorption fine structure (EXAFS) analysis revealed a strong correlation between visible light induced NIRPL and Cr-O bond distances, that was in turn inversely related to the defect concentration around Cr$^{3+}$.\cite{13} Presence of antisite defects around Cr$^{3+}$ ion increased hybridization of Cr 3d t$_{2g}$ orbitals and O 2p orbitals, resulting in shorter Cr-O bond lengths.\cite{13} In addition, X-ray absorption near edge structure (XANES) analysis showed the presence of Cr clusters (i.e. Cr$^{3+}$ ions with Cr$^{3+}$ neighbours in Ga sites) to be detrimental for NIRPL induced by visible light excitation in these spinels.\cite{14} A linear correlation was obtained between the amount of Cr clustering and the NIRPL intensity.\cite{14} The studies showed that, ZnGa$_{2}$O$_{4}$ compound with lower percentage of antisite defects and Cr clusters presented stronger persistent luminescence than MgGa$_{2}$O$_{4}$ compound with higher antisite defects and Cr clusters.\cite{14} Furthermore in a recent work by De Vos et al., it was established by first-principle calculations on ZnGa$_{2}$O$_{4}$:Cr$^{3+}$ that, the antisite defects rationalize attractive interactions between the various elements and in particular the defect energies point out the stability of two antisite defects in the vicinity of the Cr$^{3+}$ cations in ZnGa$_{2}$O$_{4}$:Cr$^{3+}$.\cite{16}

To summarize, previous investigations not only pointed out to the major role of Cr$^{3+}$ neighbouring antisite defects in inducing visible light excited NIRPL in the above mentioned spinels, but also established the key role played by Cr clusters in the quenching of NIRPL emission intensity. In this paper, we have synthesized and characterized the solid solutions of Cr$^{3+}$ doped ZnGa$_{2}$O$_{4}$ and MgGa$_{2}$O$_{4}$ mixtures in different proportions, in order to control the amount of inversion defects and to validate our earlier finding that the presence of Cr cluster defects is detrimental to NIRPL emission. Optical absorption, X-ray absorption and EPR measurements have been carried out on compounds made by solid state synthesis and the results are compared with the standard Zn$_{0.99}$Ga$_{1.99}$Cr$_{0.01}$O$_4$ (referred to as d-ZGO) sample which showed highest NIRPL\cite{2,12,17} and the other end member Mg$_{0.99}$Ga$_{1.99}$Cr$_{0.01}$O$_4$ (referred to as d-MGO) sample\cite{9,13,14}. The results once again indicate that the formation of Cr clusters leads to quenching of NIRPL in these samples. Both spinels are wide band gap semiconductors and belong to cubic space group \textit{Fd3m} with lattice parameter \textit{a} = 8.334 \AA ~for ZnGa$_{2}$O$_{4}$\cite{18} (noted as ZGO) and \textit{a} = 8.2891 \AA ~for MgGa$_{2}$O$_{4}$\cite{19} (noted as MGO). Zinc gallate compound crystallizes in normal spinel structure with Zn$^{2+}$ ions in tetrahedral coordination and Ga$^{3+}$ ions in octahedral coordination with a small cationic inversion.\cite{20,21,22} On the other hand, magnesium gallate is reported to be an almost inverse spinel with nearly 44\% octahedral site inversion.\cite{19,9}

\section{Experimental}

The samples were prepared via solid state method with their respective metal oxides ZnO (Sigma-Aldrich 99.99\% pure), Ga$_2$O$_3$ (Sigma-Aldrich 99.99\% pure), MgO (Sigma-Aldrich 99.995\% pure) and CrO$_3$ (SRL 99\% pure) as precursors. Weighed powders were thoroughly mixed along with propan-2-ol in an agate mortar. Dried mixture was pelletized in a hydraulic press under 4 tons pressure and the pellets were annealed in air at 1300$^\circ$C for 6 hours. Naturally cooled pellets were crushed into fine powders for further characterization. Zn$_{1-x}$Mg$_x$Ga$_{1.99}$Cr$_{0.01}$O$_4$ (noted ZMGO here after) compounds were prepared in four different molar ratios with x=0.1 (noted ZMGO.Mg0.1), x=0.2 (noted ZMGO.Mg0.2), x=0.3 (noted ZMGO.Mg0.3) and x=0.5 (noted ZMGO.Mg0.5).

A Rigaku X-ray diffractometer was used to obtain the X-ray diffraction (XRD) patterns at RT using Cu-K$\alpha$ radiation. The spectra were recorded in 2$\theta$ range 20$^\circ$-80$^\circ$ with 0.02$^\circ$ step and 2$^\circ$/min scan speed. Rietveld refinement on the XRD patterns was carried out using FullProf software.\cite{23}

Room temperature (RT) photoluminescence (PL) excitation spectra were recorded on Varian Cary Eclipse spectrofluorimeter in the range 190 nm-650 nm with xenon lamp (150 W) as excitation source. Pulsed laser excited PL emission measurements were run on 8 mm-diameter pellets silver glued on the cold finger of a cryogenic system maintained at 20 K. The emitted light was collected by an optical fiber and transmitted to a ICCD PI-max4 Princeton Instrument camera coupled to a Princeton Instrument Acton SpectraPro 2300 monochromator with 1200 groves/mm grating. The pellets were excited at 230 nm by an optical parametric oscillator (OPO) EKSPLA NT342B. NIRPL measurements were carried out at RT on samples weighing 180 mg filled into a 1 cm diameter circular sample holder. The samples were illuminated for 15 minutes with X-rays (Mo-tube, 20 mA-50 kV) and after the end of excitation, emission was collected using a Scientific Pixis 100 CCD camera via an optical fiber linked to an Acton SpectraPro 2150i spectrometer for spectral analysis. X-rays were used to excite the sample rather than ultra-violet (UV) source, in order to avoid variation in UV excitation wavelength due to changing band gap of the samples with increasing Mg concentration. NIRPL emission spectra were recorded during the excitation and 5 s after the end of excitation. All the samples were bleached at 250$^\circ$C for 30 minutes and were kept in the dark prior to persistent luminescence measurements.

X-ray absorption fine structure (XAFS) spectra at RT were measured on the samples in fluorescence mode for Cr K edge at BL-9A beamline in Photon Factory, Japan. Water cooled Si (111) double crystal was used as the monochromator. Absorbers were made by spreading fine powder of the samples (less than 100 microns particle size) on a scotch tape, avoiding any sort of sample inhomogeneity and pin holes. Fluorescence yield was collected via 35 pixels Ge SSD detector. EXAFS fitting was carried out using Ifeffit software with Athena and Artemis programs.\cite{24} EXAFS data for Cr K edge were Fourier transformed in the k range of 3 to 10 \AA$^{-1}$ and the fitting was performed in the R range of 1 to 3.3 \AA ~to obtain reasonable fits. Theoretical amplitude and phase information for various scattering paths were obtained using FEFF 6.01\cite{25} and the Rietveld refined parameters.

XANES spectra were calculated using FEFF 8.4 software based on the self-consistent real-space multiple-scattering formalism.\cite{26} These calculations take into account multiple scattering of photoelectron from neighbouring atoms in presence of a fully relaxed core hole. Atomic coordinates were generated for the respective ideal spinel structure with their Rietveld refined lattice parameters, using ATOMS 2.5 version to obtain a FEFF output file.\cite{27} The Ga core in the FEFF file was changed to Cr before running the FEFF program, to calculate the XANES pattern over a radius of about 8 \AA. The defects were then introduced accordingly around Cr in respective coordinates discussed later in the paper, and the spectra were computed for each case. These calculated spectra were linearly combined to obtain a best fit for the experimental spectra.

EPR spectra were recorded at room temperature with a Bruker Elexsys E500 spectrometer operating at X-band ($\approx$ 9.4 GHz) in continuous-wave mode and equipped with a Bruker 4122SHQE/0111 resonator. The EPR spectra of powders were recorded with a magnetic field modulation at 100 kHz and with amplitude 0.5 mT.

\section{Results}

\begin{figure}[h]
\centering
\includegraphics[width=\columnwidth]{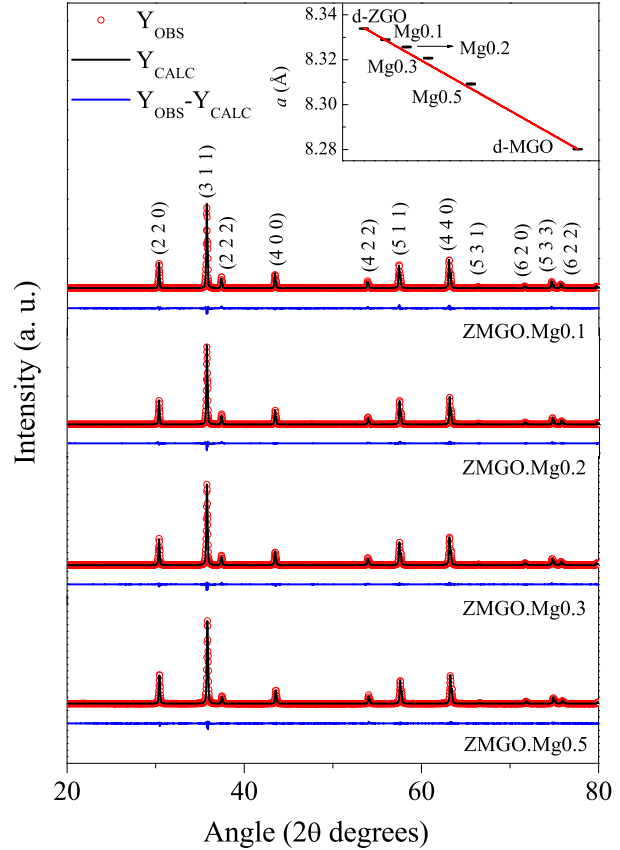}
\caption{XRD patterns of ZMGO compounds along with their Rietveld refined fits and residual patterns; Inset shows the corresponding lattice parameters obtained from Rietveld refinement.}
\label{fig1}
\end{figure}

X-ray diffraction measurements were carried out on all the Cr$^{3+}$ doped ZMGO compounds. The XRD patterns of these compounds along with their corresponding Rietveld refined patterns and residual patterns are presented in Figure \ref{fig1}. As evident from Figure \ref{fig1}, XRD indicates the formation of phase pure cubic spinel compounds. However as will be discussed in the later sections, EPR shows the presence of a very small amount (less than 1\%) of magnesium oxide MgO. The lattice constants obtained from the Rietveld refinement along with their errors are plotted as a function of Mg concentration, in the inset of Figure \ref{fig1}. The lattice parameters of Cr$^{3+}$ doped d-ZGO and d-MGO compounds\cite{13,14} are shown for reference. It can be visibly noted from the graph that, the lattice parameter \textit{a} is linearly decreasing with the increase in Mg concentration. This is expected as the ionic size of Mg$^{2+}$ ion both in tetrahedral and octahedral coordination is smaller than that of Zn$^{2+}$ ion, leading to a decrease in \textit{a}. The refinement of cationic site occupancy indicated increasing structural inversion, showing increase in Mg$^{2+}$ ions occupancy of octahedral sites, with increasing Mg concentration. The Mg occupancy in octahedral sites was found to be 3.1(4)\% for ZMGO.Mg0.1, 7.0(3)\% for ZMGO.Mg0.2, 11.8(3)\% for ZMGO.Mg0.3 and 21.1(3)\% for ZMGO.Mg0.5 compound.

\begin{figure}[h]
\centering
\includegraphics[width=\columnwidth]{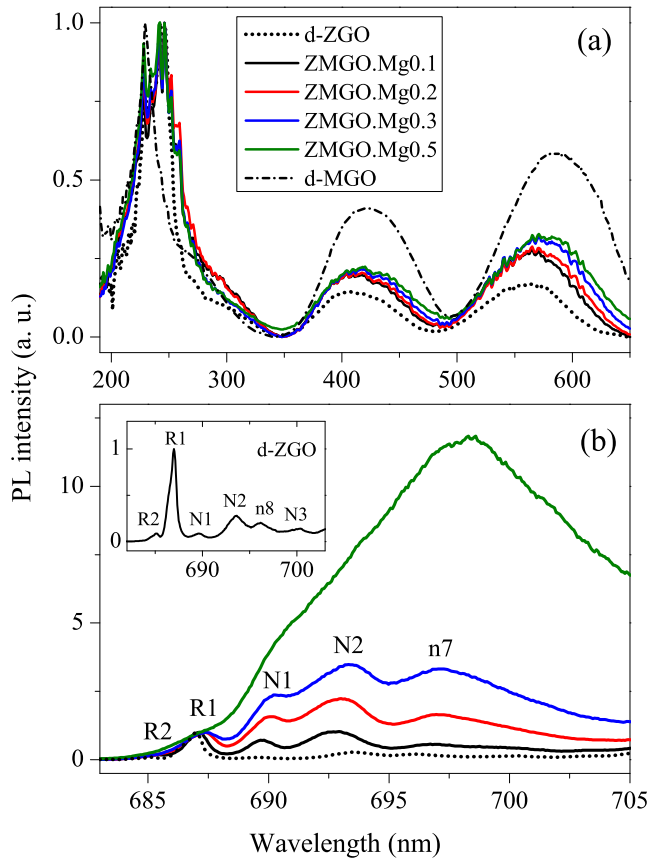}
\caption{(a) PL excitation spectra of Cr$^{3+}$ doped ZMGO compounds with varying Mg concentration, d-MGO and d-ZGO compounds measured at RT. (b) Corresponding Zero-phonon PL emission spectra at 20 K and excited at 230 nm; Inset shows the low temperature PL spectrum of d-ZGO for better comparison.}
\label{fig2}
\end{figure}

PL excitation spectra measured at RT for Cr$^{3+}$ doped ZMGO compounds along with d-ZGO and d-MGO PL excitation spectra are presented in Figure \ref{fig2}(a). The spectra consist of host band gap excitation peaking at about 245 nm for all the compounds and three broad absorption bands at around 290 nm, 415 nm and 565 nm belonging to $^4$A$_2$($^4$F) $ \rightarrow $ $^4$T$_1$($^4$P), $^4$A$_2$($^4$F) $ \rightarrow $ $^4$T$_1$($^4$F) and $^4$A$_2$($^4$F) $ \rightarrow $ $^4$T$_2$($^4$F) Cr$^{3+}$ d-d transitions respectively.\cite{28,29,30} Intensity of the Cr absorption bands is observed to increase with increasing Mg substitution and the bands are seen to be gradually moving towards longer wavelengths. It must be noted here that, the red shift seen in Cr$^{3+}$ absorption bands is due to the weaker crystal field around Cr$^{3+}$ ion in MGO than that around Cr$^{3+}$ ion in d-ZGO,\cite{9,31} which is due to the presence of comparatively higher amount of defects around Cr$^{3+}$ ion in MGO\cite{9}. The increase of d-d transition intensity upon increasing magnesium content is likely due to a progressive lowering of Cr site symmetry, which increases the oscillator strength of these bands.\cite{29}

PL emission spectra of Cr$^{3+}$ doped ZMGO and d-ZGO compounds measured at 20 K with 230 nm excitation, corresponding to $^2$E($^2$G) $ \rightarrow $ $^4$A$_2$($^4$F) Cr$^{3+}$ d-d transition\cite{28,29} are presented in Figure \ref{fig2}(b) with their intensities normalized to R1 line. The spectrum corresponding to d-ZGO compound (shown in inset of Figure \ref{fig2}(b)) displays six zero phonon lines (ZPL) – R2, R1, N1, N2, n8 and N3, whose origins are explained in our previous papers\cite{12,13}. N1 and N2 lines attributed to Cr$^{3+}$ close to Ga$_{Zn}$ antisite defects\cite{8} and to pairs of antisite defects Ga$_{Zn}$ and Zn$_{Ga}$, respectively\cite{32,33,34,12}, are seen to grow proportionally with increasing Mg concentration in ZGO. Also, their peak maximum shifts to longer wavelengths indicating weakening of crystal field around Cr$^{3+}$ due to the presence of these defects. The n7 line arising due to the formation of Cr$^{3+}$ clusters\cite{12} is seen to be rapidly growing with the increase in Mg substitution in the samples, masking the n8 and N3 lines which were seen in d-ZGO. In the ZMGO.Mg0.5 compound spectrum, which is very similar to the completely inverse spinel MGO compound spectrum reported earlier\cite{9}, the emission lines are hardly distinguishable due to their overlapping nature and the large increase in n7 line intensity. However, one can say that there is a gradual change of structure with Mg doping, from ZGO-like structure to MGO-like structure.

\begin{figure}[h]
\centering
\includegraphics[width=\columnwidth]{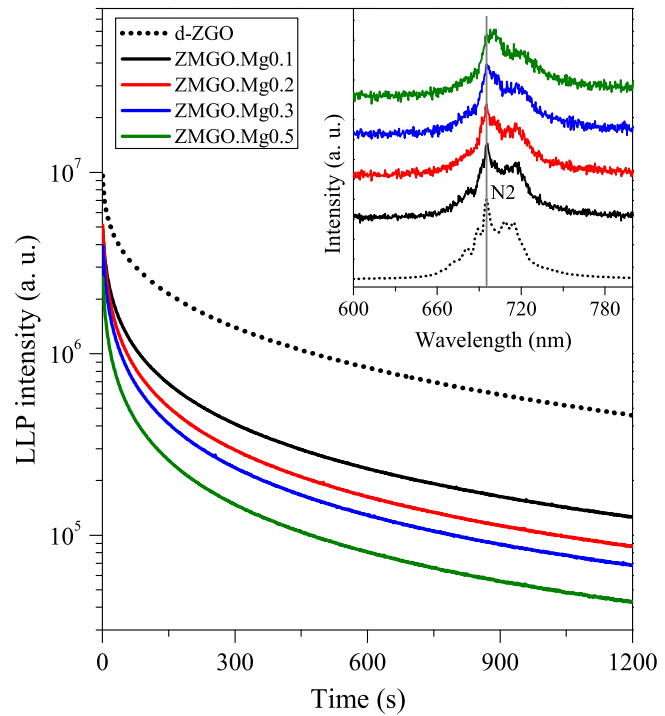}
\caption{X-rays irradiated NIRPL decay curves measured for ZMGO compounds along with that for d-ZGO compound for reference. Inset shows the corresponding emission spectra for the NIRPL.}
\label{fig3}
\end{figure}

NIRPL intensity decay curves of ZMGO compounds along with d-ZGO, recorded after 15 minutes of X-ray illumination are plotted in Figure \ref{fig3}. The increase in the Mg concentration associated with increase in N-line intensities had a direct effect on the NIRPL intensity of these compounds. ZMGO.Mg0.1 compound showed highest NIRPL intensity and ZMGO.Mg0.5 compound showed the lowest at the tail part. The compounds were seen to present less NIRPL compared to d-ZGO compound. A huge decrease in intensity with just 10\% Mg in the lattice could be ascribed to the presence of large amounts of Cr$^{3+}$ clusters which are previously shown to be detrimental for NIRPL.\cite{14} The NIRPL emission spectra are shown in the inset of Figure \ref{fig3}. The ZMGO spectra show dominant N2 emission line, similar to d-ZGO.\cite{2,12,13,35} But this N2 line is seen to be shifting to longer wavelengths and the emission spectra broadens (Figure \ref{fig3} (inset)) with increase in Mg concentration.

\begin{figure}[h]
\centering
\includegraphics[width=\columnwidth]{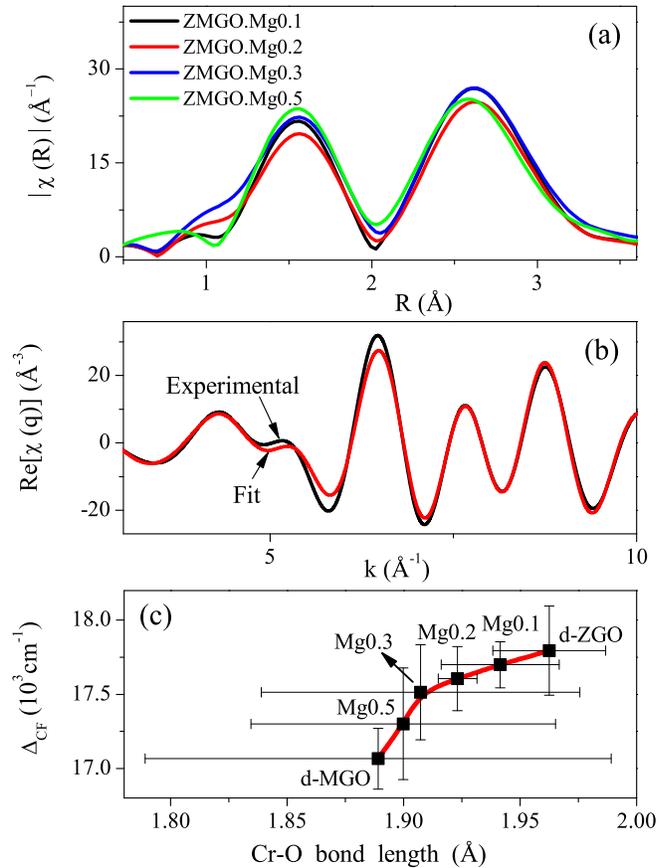}
\caption{(a) Cr K edge FT magnitude of EXAFS patterns for ZMGO compounds, Fourier transformed in the k range 3 to 10 \AA$^{-1}$. (b) Experimentally obtained Cr K edge EXAFS pattern for ZMGO.Mg0.1 compound along with fit in q space in the k range 3 to 10 \AA$^{-1}$. (c) Cr-O bond lengths obtained from EXAFS fitting plotted versus crystal field energy $\Delta_{CF}$ deduced from $^4$A$_2$ $ \rightarrow $ $^4$T$_2$ Cr$^{3+}$ absorption band in RT PL spectra.}
\label{fig4}
\end{figure}

X-ray absorption studies were carried out on Cr$^{3+}$ doped ZMGO compounds to complement the optical studies. XAFS measurements were performed at Cr K edges in fluorescence mode on all the four ZMGO compounds. Magnitude of Fourier transform (FT) of Cr K edge EXAFS for these compounds are presented in Figure \ref{fig4}(a). The first peak corresponding to Cr-O correlation is seen to be shifting to lower R values with more Mg substitution in ZGO sample. This implies that the average Cr-O distance is decreasing and indicates an increase in distortions in Cr$^{3+}$ local environment. The peaks were observed to broaden for higher Mg concentrations, again indicating the distorted local octahedral environment around Cr$^{3+}$ ion.

Experimentally obtained EXAFS spectra were fit for all the ZMGO compounds using Ifeffit software. The spectra were computed using two structural models based on normal spinel ZGO structure with Zn occupying tetrahedral sites and Ga occupying octahedral sites; and inverse spinel MGO model wherein the Mg ions occupied octahedral sites and half of the Ga ions occupied the tetrahedral sites. To elaborate, the second peak of the spectra were fit using Cr-Ga and Cr-Zn paths from ZGO spinel model and, Cr-Mg and Cr-Ga paths from MGO inverse spinel model.  Phase fraction of each model was taken as a fitting parameter and the contribution of inverse spinel model gave the amount of inversion. However, the first peak of the spectra were fit using a single Cr-O path, since the number of free parameters to vary were limited by the available data (k$_{max}$ = 10 \AA$^{-1}$).  An example of fit to EXAFS data in back transformed \textit{k} space is presented in Figure \ref{fig4}(b) for ZMGO.Mg0.1 compound.

Cr-O bond lengths obtained from the Cr k edge EXAFS fitting of ZMGO, d-ZGO and d-MGO compounds are plotted versus crystal field energy $\Delta_{CF}$ in Figure \ref{fig4}(c). $\Delta_{CF}$ for each compound is calculated from $^4$A$_2$ $ \rightarrow $ $^4$T$_2$ Cr$^{3+}$ absorption band in their corresponding PL excitation spectrum (Figure \ref{fig2}(a)). As seen from the Figure \ref{fig4}(c), both Cr-O bond lengths and $\Delta_{CF}$ are decreasing with increase in Mg substitution in ZGO. We have reported in our previous paper that, increase in defect concentration around Cr$^{3+}$ results in shortening of Cr-O bond lengths and lowering of crystal field energies, due to the increasing $\pi$ bond contribution to the Cr-O interaction.\cite{13} Thus, the observed lowering of Cr-O bond distances and $\Delta_{CF}$ in ZMGO compounds for higher Mg concentrations (Figure \ref{fig4}(c)) can be attributed to the increasing defect concentration, in consistence with the optical results. We can also see from Figure \ref{fig4}(c) that the error bars for Cr-O bond lengths are large. Since a single Cr-O path was used in our fitting as mentioned in previous paragraph, a distribution of Cr-O distances\cite{11,13} (which leads to uneven broadening of the first peak as seen from Figure \ref{fig4}(a)) is unaccounted in the fitting. Hence, this distribution of Cr-O distances manifests as large error bar in Cr-O bond lengths.

\begin{figure}[h]
\centering
\includegraphics[width=\columnwidth]{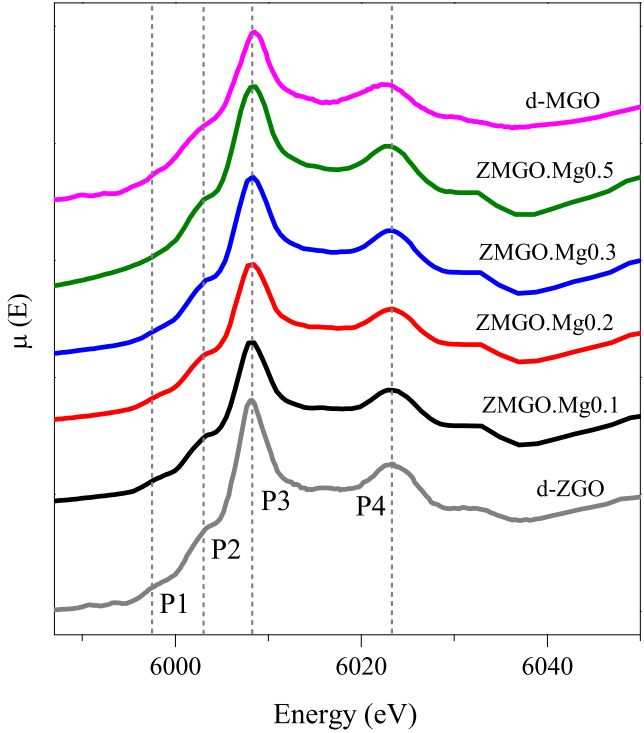}
\caption{Experimental XANES spectra of all the four ZMGO compounds and the end members d-ZGO and d-MGO.}
\label{fig5}
\end{figure}

Experimentally obtained Cr K edge X-ray absorption near edge structure (XANES) spectra recorded for all the four Cr$^{3+}$ doped ZMGO compounds are presented in Figure \ref{fig5}, along with that of d-ZGO and d-MGO compounds for reference. Each spectrum shows the four prominent features – P1, P2, P3 and P4 (marked by dotted lines for d-ZGO) in the near edge region.

\begin{figure}[h]
\centering
\includegraphics[width=\columnwidth]{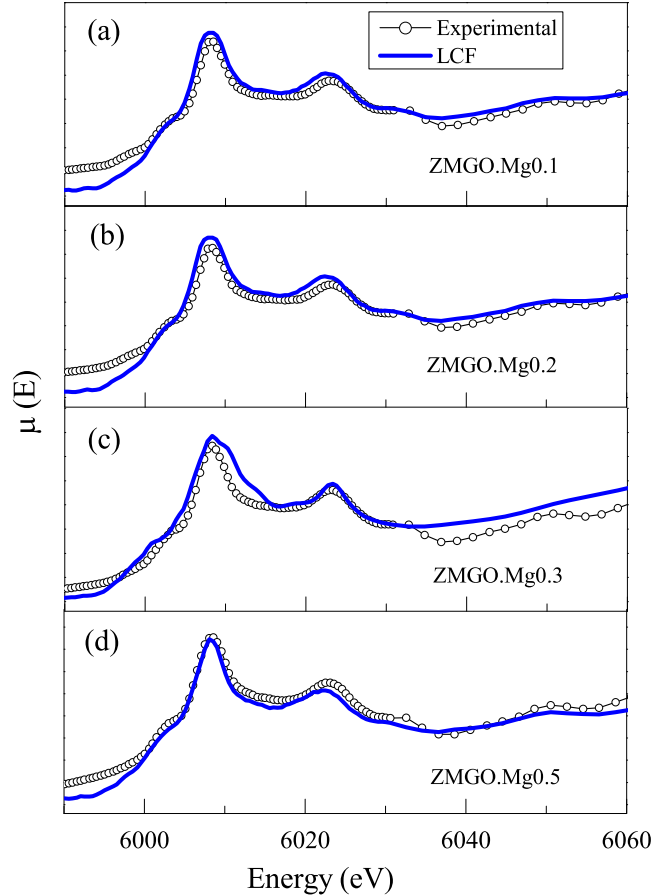}
\caption{Experimental Cr-K edge XANES spectrum along with computed LCF for (a) ZMGO.Mg0.1, (b) ZMGO.Mg0.2, (c) ZMGO.Mg0.3 and (d) ZMGO.Mg0.5.}
\label{fig6}
\end{figure}

In our previous paper\cite{14}, ab-initio XANES spectra were computed for ZGO and MGO compounds with various defect scenarios like Cr in ideal octahedral environment with Cr-O distance of 2 \AA ~(ideal), Cr with antisite defect in its neighbourhood with Cr-O distance of 1.9 \AA ~(antisite) and Cr with Cr ion as its first cationic neighbour with Cr-O distance of 2\AA ~(Cr clusters), to calculate the amount of defects around Cr$^{3+}$. In order to estimate the contribution of each of the calculated spectra to the experimental curve, a linear combination fitting (LCF) was performed of all possible combinations of the calculated spectra using Athena software for XAFS analysis.\cite{14} Here, the LCF was performed on the experimentally obtained d-ZGO and d-MGO spectra\cite{14}, to obtain best fitted (lowest R factor) curves in case of ZMGO compounds. The obtained best fits along with the experimental spectra are presented in Figure \ref{fig6}. The combinations used here for the fitting, which include experimentally obtained d-ZGO spectrum and d-MGO spectrum, are already a combination of ideal spectrum with Cr-O distance $\approx$ 2 \AA, 6 neighbouring antisite defects spectrum with Cr-O distance $\approx$ 1.9 \AA ~and Cr$^{3+}$ with Cr as its first cationic neighbour (referred to as Cr clusters) spectrum with Cr-O distance $\approx$ 2 \AA.\cite{14} The percentages of ideal, antisites and Cr clusters spectra contributions to the ZMGO spectra were calculated from the obtained d-ZGO and d-MGO combinations. The derived percentages for each compound are tabulated in Table \ref{table1}.

\begin{table}[t]
\centering
\caption{Percentages of different spectra obtained from LCF of ZMGO compounds, along with d-ZGO and d-MGO compounds for reference. Typical errors in the values are of the order of 5\%.}
\label{table1}
\begin{tabular*}{\linewidth}{l c c c}
\hline
Samples & Ideal & Antisites & Cr clusters  \\ [0.5ex] \hline
d-ZGO & 71.2 & 16.4 & 12.4 \\ [0.5ex]
ZMGO.Mg0.1 & 42.0 & 18.9 & 39.1 \\ [0.5ex]
ZMGO.Mg0.2 & 41.4 & 18.9 & 39.6 \\ [0.5ex]
ZMGO.Mg0.3 & 30.7 & 21.9 & 47.4 \\ [0.5ex]
ZMGO.Mg0.5 & 9.5 & 21.6 & 68.9 \\ [0.5ex]
d-MGO & 9.5 & 21.6 & 68.9 \\ [0.5ex]
\hline
\end{tabular*}
\end{table}

In order to test the validity of the fitting procedure used in XANES, EPR spectra of ZMGO compounds were recorded and fitted with linear combinations of experimental spectra of d-ZGO and d-MGO compounds. With transition linewidths generally well below 10$^{-5}$ eV (compared to $\sim$ 5-10 eV for XANES), EPR is sensitive to minute variations in the crystal field around the ion.

\begin{figure}[h]
\centering
\includegraphics[scale=0.5]{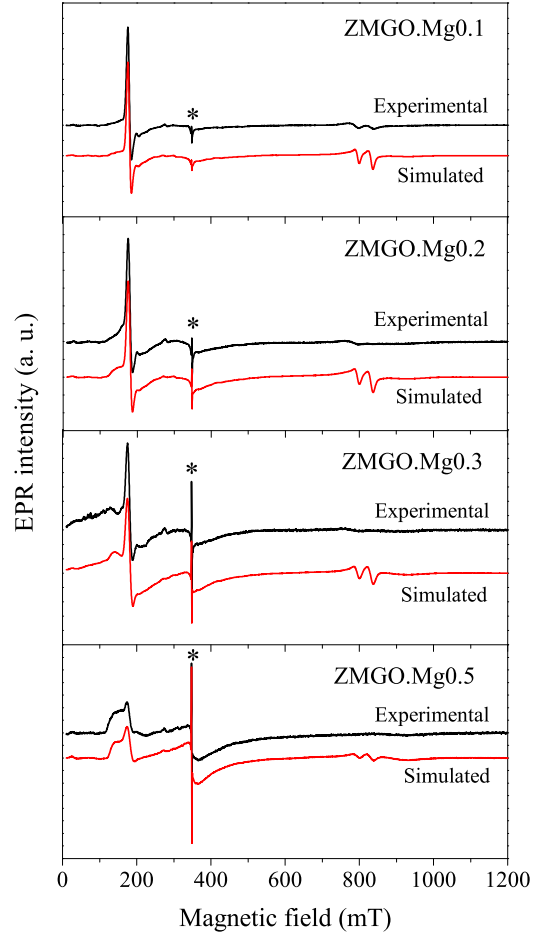}
\caption{X-band EPR spectra at room temperature of Cr$^{3+}$ in ZMGO compounds. They are fitted by linear combinations of experimental spectra of d-ZGO and d-MGO compounds, with the additional contribution from a broad component, and a very narrow line due to Cr$^{3+}$ in magnesium oxide MgO impurities (marked by a star).}
\label{fig7}
\end{figure}

\begin{figure}[h]
\centering
\includegraphics[width=\columnwidth]{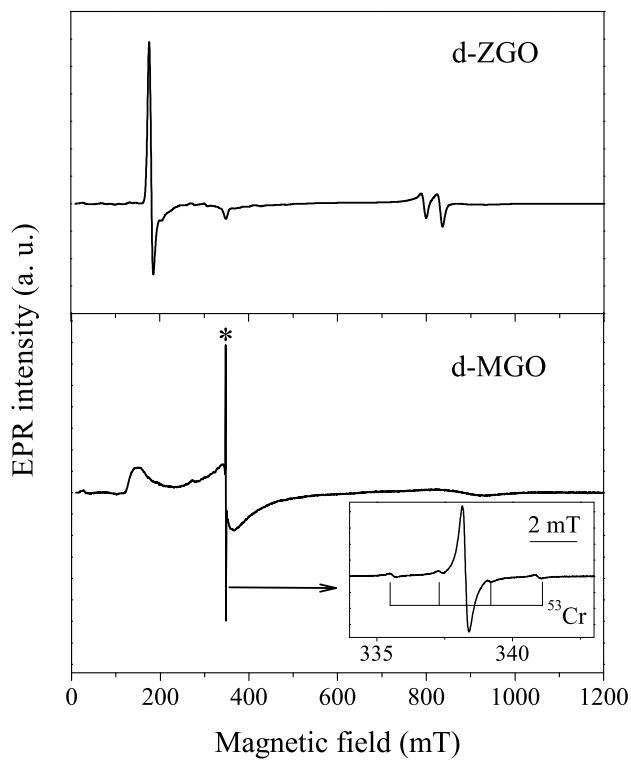}
\caption{X-band EPR spectra at room temperature of Cr$^{3+}$ in d-ZGO and d-MGO compounds. The very narrow line around 350 mT (marked by a star) is due to Cr$^{3+}$ in magnesium oxide MgO impurities. The inset represents a zoom of this narrow line, showing the satellites due to hyperfine interaction with $^{53}$Cr isotope.}
\label{fig8}
\end{figure}

\begin{figure}[h]
\centering
\includegraphics[width=\columnwidth]{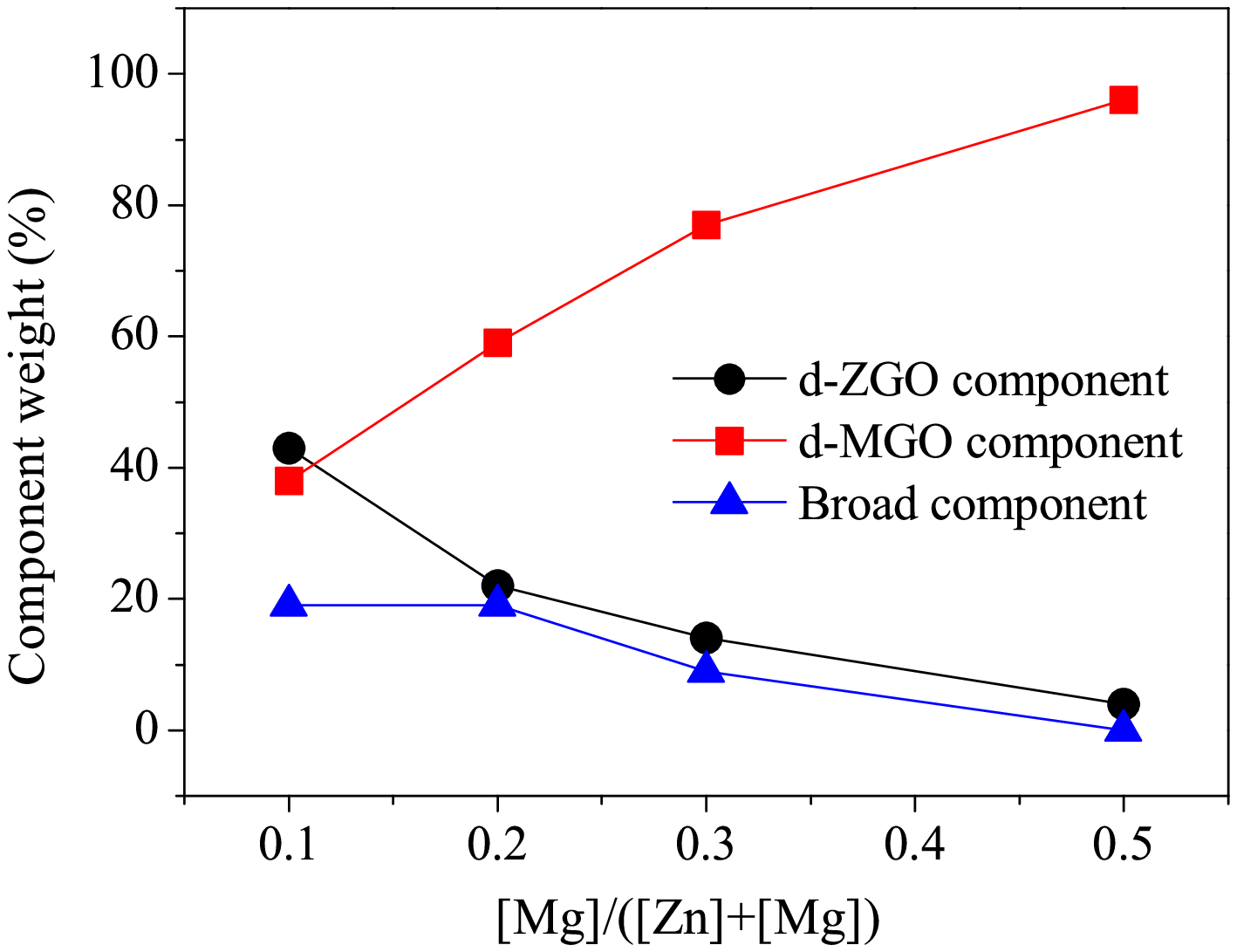}
\caption{Variation with Mg content of ZMGO of the two types of environments of Cr$^{3+}$, deduced from EPR spectra. The broad component was necessary to reproduce the spectra for low Mg contents.}
\label{fig9}
\end{figure}

X-band EPR spectra of the Cr$^{3+}$ in ZMGO compounds with various Mg concentrations, and in pure d-ZGO and d-MGO compounds are shown in Figures \ref{fig7} and \ref{fig8}, respectively. The narrow line indicated by a star (at 350 mT) arises from Cr$^{3+}$ impurities in pure magnesium oxide (MgO) precipitates which are present in amounts below the XRD detection threshold (probably less than 1\%). This interpretation of the narrow line is argued by the observation of hyperfine satellites due to $^{53}$Cr isotope (nuclear spin \textit{I}=3/2, natural abundance 9.5\%) (see the inset in Figure \ref{fig8}), and the observed very small linewidth and the lack of zero-field splitting of this line prove that this Cr$^{3+}$ impurity is located in a purely cubic phase, which is the case of MgO. Thus, this impurity will no longer be discussed further. It must just be noted that its intensity increases with the amount of Mg content in the compounds, indicating an increasing amount of magnesium oxide. When compared to the spectra of Cr$^{3+}$ in d-ZGO and in d-MGO (Figure \ref{fig8}), the spectra of the ZMGO samples look like a combination of at least these two spectra. Therefore, in a similar approach as for the simulation of the XANES spectra, we simulated the EPR spectra with a linear combination of 3 basis components: (i) a spectrum of Cr$^{3+}$ in d-ZGO corresponding mostly to Cr$^{3+}$ in the ideal site of ZGO i.e. with D$_{3d}$ symmetry, (ii) a spectrum of Cr$^{3+}$ in d-MGO corresponding to a Cr$^{3+}$ in a highly distorted site (no axial symmetry) due to a large number of neighbouring antisites defects and neighbouring Cr$^{3+}$ ions, (iii) a broad component represented either by a spectrum similar to that of d-ZGO but with a much larger linewidth (case of ZMGO.Mg0.1 and ZMGO.Mg0.2) or a broad lorentzian line centered at 190 mT for ZMGO.Mg0.3. This third component was however not necessary for ZMGO.Mg0.5 which presented a spectrum very similar to that of d-MGO. The broad component could possibly arise from Cr$^{3+}$ ions with high local concentration and thus experiencing strong dipolar interactions. It is the equivalent of clusters observed by XANES. However, in addition to these dipolar interactions, antiferromagnetic interactions between neighbouring Cr$^{3+}$ ions may strongly reduce the intensity of this component or eventually give EPR silent clusters, so that our simulations cannot give information on the amount of Cr$^{3+}$ in clusters. Figure \ref{fig7} shows the simulated spectra obtained by combining the three basis components with relative weights determined by least-square fitting. The agreement between the simulated and the experimental spectra is fairly good in the low field part (field $<$ 600 mT). There is a slight discrepancy concerning the signals in the high field part of the spectrum at about 800 mT. This pair of lines, which arises from the d-ZGO type component, is very sensitive to broadening induced by crystal field disorder\cite{12} (strain broadening), while crystal field disorder does not affect the low field part of the d-ZGO spectrum. This strain broadening is weak in the basis d-ZGO spectrum used for decomposition while it is stronger in the actual ZGO type component of the ZMGO spectra. Therefore, the low field part of the spectra is more relevant for assessing the agreement between experimental and simulated spectra. The proportions of each component in the ZMGO spectra were determined from their simulation with the Easyspin software and double integration of the simulated spectra. These proportions are reported in Figure \ref{fig9}.

One might logically imagine that going from pure ZGO to pure MGO by Mg/Zn substitution would induce a progressive modification of the environment around each Cr$^{3+}$. Instead of that, these simulations of XANES and EPR spectra indicate that increasing the Mg substitution in the ZMGO series maintain the coexistence of two types of Cr$^{3+}$ environments, namely Cr$^{3+}$ having a predominantly ordered environment (ZGO type environment) and Cr$^{3+}$ perturbed by the presence of a large amount of neighbouring antisite defects and neighbouring Cr$^{3+}$ ions (MGO-type environment).

From Table \ref{table1} (XANES) and also from Figure \ref{fig9} (EPR), one can observe that from ZGO to MGO, contribution from Cr$^{3+}$ in undistorted sites is decreasing, whereas contributions from Cr$^{3+}$ (Figure \ref{fig2}(b)) wherein N2 and n7 emission line intensities corresponding to antisite and Cr cluster defects respectively are increasing for higher Mg concentrations. The percentage of Cr clusters is increasing and the ideal spectrum percentage is decreasing in ZMGO.Mg0.1 as compared to ZGO which results in the observed lowering of NIRPL intensity (Figure \ref{fig3}). From ZMGO.Mg0.1 to ZMGO.Mg0.2, there is a slight decrease in ideal spectrum contribution and slight increase in Cr clusters contribution which results in a slightly lower NIRPL intensity for ZMGO.Mg0.2 compound (Figure \ref{fig3}). The ideal spectrum percentage then decreases sharply concomitant with equally sharp increase in percentage of Cr clusters for ZMGO.Mg0.3 compound. This is also reflected in lower NIRPL intensity for ZMGO.Mg0.3 as compared to ZMGO.Mg0.2. The LCF fit to the XANES spectrum of ZMGO.Mg0.5 compound resulted in identical percentages to that of d-MGO spectrum, showing a steep decrease in ideal component contribution and an increase in Cr clusters contribution, and a corresponding decrease of NIRPL intensity. This result clearly substantiates our previous hypothesis that formation of Cr$^{3+}$ clusters leads to quenching of NIRPL.\cite{14}

\begin{figure}[h]
\centering
\includegraphics[width=\columnwidth]{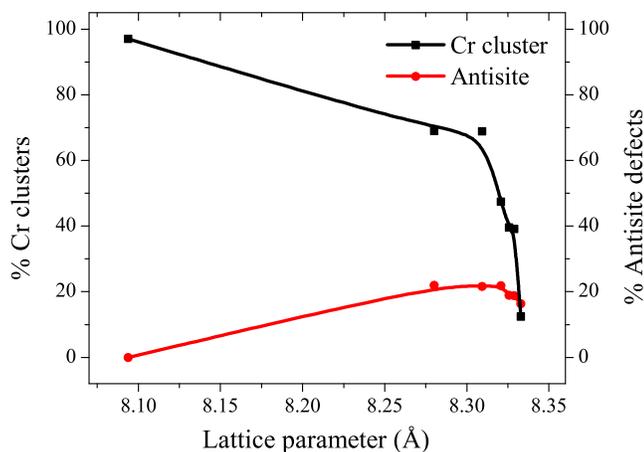}
\caption{A correlation between lattice constants and percentages of Cr$^{3+}$ clusters and  of Cr$^{3+}$ with antisite defects in the spinels, deduced from XANES spectra.}
\label{fig10}
\end{figure}

As previously reported, the lattice parameter increases from ZAO (zinc aluminate) to MGO to ZGO.\cite{13} On the contrary, the \% of Cr clusters increases in the reverse order i.e. ZGO to MGO to ZAO.\cite{13} This could be an indication of the correlation between lattice parameters and the formation of antisite defects and Cr clusters. The lattice constant for ZnCr$_2$O$_4$ normal spinel compound, wherein Cr$^{3+}$ occupies octahedral position, is reported to be \textit{a} $\approx$ 8.33 \AA \cite{13} which is very close to the lattice parameter of ZGO (\textit{a} $\approx$ 8.33 \AA)\cite{18}. Since lattice parameter decreases for other two compounds (MGO and ZAO), there is a reduction in their unit cell volume. However, when Cr$^{3+}$ is doped in these compounds, it tries to modify its surroundings, similar to its octahedral environment in ZnCr$_2$O$_4$. The higher the difference in lattice parameter between ZnCr$_2$O$_4$ and the spinel compound with Cr ion, the higher is the percentage of Cr clusters formation. This is irrespective of whether the host lattice is a normal spinel or supports cationic site inversion. This is presented in Figure \ref{fig10} wherein the percentage of Cr clusters is seen to increase with a decrease in lattice parameter. Figure \ref{fig10} also shows the variation of antisite defect percentage as a function of lattice parameter. Though, with decrease in lattice parameter, there is a small increase in the antisite defect contribution in going from ZGO to MGO, the increase in Cr cluster contribution is much more. This leads us to believe that the decrease in lattice parameter is the driving force for cluster formation. This needs to be further confirmed by some first principles calculations. To summarize, the possibility and the amount of formation of Cr clusters increases when the lattice parameter of the spinel host decreases as compared to that of ZnCr$_2$O$_4$ compound.

\section{Conclusions}

A series of ZGO compounds prepared with Mg substituting Zn showed an increase in antisite defects with increasing Mg concentration, from optical, EPR and structural characterizations. Photoluminescence and X-ray absorption studies showed an increase in the amount of Cr clusters in these compounds with greater Mg substitution, which resulted in the corresponding decrease of NIRPL intensity, substantiating our hypothesis that Cr clusters are detrimental to persistent luminescence. The possibility of formation of Cr clusters was shown to be inversely related to the lattice parameter of the spinel hosts.

\section{Acknowledgment}

This work was done as a part of Photon Factory beamtime proposal 2012G559. Financial support from Indo-French Centre for the Promotion of Advanced Research (IFCPAR)/ CEntre Franco-Indien Pour la Recherche Avanc\'{e}e (CEFIPRA) is acknowledged.

\end{document}